# Um Sistema Multiagente no Combate ao Branqueamento de Capitais

Claudio Alexandre, João Balsa

**calexandre@di.fc.ul.pt, jbalsa@ciencias.ulisboa.pt.**

Faculdade de Ciências da Universidade de Lisboa, BioISI-MAS, Campo Grande, 1749-016, Lisboa, Portugal

**Resumo:** Branqueamento de capitais é um crime que possibilita o financiamento de outros crimes, por isso ele é importante para as organizações criminosas e seu combate é motivo de mobilização das nações do mundo inteiro. O processo de anti-branqueamento de capitais não evoluiu como esperado pois tem priorizado a sinalização de transações suspeitas. O crescente aumento no volume de transações tem sobrecarregado o indispensável trabalho humano de avaliação final das sinalizações. Este artigo apresenta um sistema multiagente que objetiva ir além da captura de transações suspeitas, buscando auxiliar o especialista humano na análise das suspeições. Os agentes criados utilizam técnicas de data mining para criação de perfis de comportamento transacional; aplicam as regras obtidas no aprendizado em conjunto com regras especificas baseadas em aspectos legais e nos perfis criados para captura de transações suspeitas; e analisam estas transações sinalizadas indicando ao especialista humano aquelas que necessitam de análise mais detalhada.

**Palavras-chave**: Sistemas multiagente; agentes inteligentes; data mining; anti-branqueamento de capitais.

### *A Multi-Agent System in the Combat Against Money Laundering*

***Abstract:*** *Money laundering is a crime that makes it possible to finance other crimes, for this reason, it is important for criminal organizations and their combat is prioritized by nations around the world. The anti-money laundering process has not evolved as expected because it has prioritized only the signaling of suspicious transactions. The constant increasing in the volume of transactions has overloaded the indispensable human work of final evaluation of the suspicions. This article presents a multiagent system that aims to go beyond the capture of suspicious transactions, seeking to assist the human expert in the analysis of suspicions. The agents created use data mining techniques to create transactional behavioral profiles; apply rules generated in learning process in conjunction with specific rules based on legal aspects and profiles created to capture suspicious transactions; and analyze these suspicious transactions indicating to the human expert those that require more detailed analysis.*

***Keywords****: multi-agent systems; intelligent agents; data mining; anti-money laundering.*

## 1. Introdução

Ações de prevenção e combate ao crime de branqueamento de capitais (BC) são priorizadas por quase todos os governos do mundo, no mínimo, no mesmo nível das grandes questões globais (Madinger, 2012). BC é tipicamente um crime que consiste em tornar lícita a origem ilícita de um determinado ganho financeiro. A estimativa global anual de BC é em torno de 2% a 5% do Produto Interno Bruto Global (UNODC, 2014). Sobre fraude financeira, somente a América do Norte estima que sua perda passe de $3,1 mil milhões em 2017 para $4,8 mil milhões em 2018 (Cser, 2017). Este volume financeiro perdido anualmente já é motivo suficiente para o assunto ser tratado com prioridade, no entanto, outro fator leva os governos a priorizarem o combate a este crime: a sua comprovada ligação com outras práticas criminosas como narcotráfico, fraude, corrupção, sequestro, terrorismo, contrabando de armas, entre outros (Schott, 2006).

As instituições financeiras, em sua maioria, já utilizam processos semi-automatizados para sinalização de transações suspeitas de BC, baseados em informações de cadastro, médias, desvios padrões e regras fixas pré-estabelecidas, geralmente oriundas de observações empíricas ou da experiência humana dos Analistas de anti-branqueamento de capitais (ABC). Porém, o crescente aumento no volume das transações realizadas, aliado à frequente publicação de novas regulamentações, nacionais e internacionais, acabam por provocar ineficiência neste processo de sinalização.

Visando tornar o processo de ABC mais ágil e eficiente, este artigo apresenta a estrutura de um sistema multiagente para suporte à tomada de decisão neste contexto. Os agentes inteligentes utilizam técnicas de data mining para a criação de perfis de comportamento transacionais históricos, analisam e sinalizam transações suspeitas e auxiliam o Analista de ABC na tomada de decisão sobre as sinalizações. O modelo BDI (*Belief, Desire, Intention*) incorporado pela metodologia de desenvolvimento adotada (*Prometheus*), permite utilizar os perfis comportamentais encontrados, bem como possibilita a implementação de regras especificas em função do risco envolvido. Estas características, dentre outros benefícios, aumentam a eficiência do processo, enfrentando o aumento no volume de transações com a redução gradativa da intervenção humana.

Este artigo segue a seguinte organização: indicação dos trabalhos relacionados com propostas de mineração de dados e arquiteturas de sistemas que buscam o ABC; apresentação da estratégia baseada no comportamento transacional dos atores para combate a este tipo de fraude; descrição da abordagem conservadora adotada no tratamento do risco envolvido; detalhamento do sistema e dos agentes definidos; apresentação dos resultados obtidos e as conclusões.

## 2. Trabalhos Relacionados

A capacidade de adaptação do *modus operandi* dos fraudadores e a falta de informação sistematizada que associe as transações suspeitas com a comprovação do crime, são obstáculos para um avanço mais rápido da automatização do processo de prevenção e combate ao crime de branqueamento de capitais (BC).

O primeiro sistema amplamente divulgado na área do anti-branqueamento de capitais (ABC) foi o FinCEN *Artificial Intelligence System* (FAIS) (Senator *et al.*, 1995), desenvolvido e utilizado pelo *Financial Crimes Enforcement Network* (FinCEN), órgão do Departamento do Tesouro dos Estados Unidos. Em seguida, a Subcomissão Permanente de Investigações do Comitê do Senado Americano para Assuntos Governamentais solicitou ao *Office of Technology Assessement* (OTA)[1] avaliação sobre a utilização de técnicas de pesquisa baseadas em inteligência artificial (IA), visando monitorizar o tráfego e transferências bancárias com o propósito de reconhecer transações suspeitas. Em seu relatório de 1995, o OTA concluiu que "o conceito original na sua formulação mais simples – o monitoramento do tráfego de transferência bancária, de forma contínua e em tempo real, usando técnicas de inteligência artificial – não é viável". Contudo, ponderou que existiam alternativas na tecnologia da informação a serem utilizadas para apoiar e reforçar a aplicação da lei contra o BC. Sugeriu, então, a utilização de técnicas tais como: *knowledge acquisition, machine learning, clustering, knowledge sharing* e *data transformation* (OTA, 1995). Em 1998, por orientação do relatório do *United States General Accounting Office* (GAO), a gestão decidiu pela não implementação de novos produtos utilizando técnicas de IA.

Curiosamente, desde então, os estudos têm direcionado propostas para utilização das técnicas recomendadas buscando a identificação de anomalias ou situações suspeitas, conforme bem demonstrado em Chandola *et al.* (2009) e Sabau (2012). Contudo, conforme será comentado na seção 2.2, são poucas as propostas de sistemas baseados em agentes inteligentes para suportar todo o processo de ABC.

**2.1. Mineração de Dados**

As técnicas de *data-mining*, *machine learning* e *clustering* têm sido fortemente utilizadas na tentativa de identificar casos suspeitos de BC. Em Zhang *et al.* (2003) um conjunto de dados são discretizados, mapeados para espaço dimensional Euclidiano, projetados numa linha de tempo discretizada, formando um histograma. Os clusters são criados, utilizando o algoritmo K-means, com base nos segmentos do histograma. Análises de correlação locais e globais são então aplicadas para detectar padrões suspeitos. É uma boa abordagem para a análise de comportamentos individuais e/ou de grupo baseado em picos anormais no histograma. No entanto, na análise de grande quantidade de clientes e transações durante longo período de tempo, a detecção de casos suspeitos pode ser dificultada, pois podem existir poucos ou nenhum pico no histograma.

Em Kingdon (2004), é proposta a utilização de técnicas de IA e de uma máquina de vetores de suporte (*Support Vector Machine*-SVM) numa perspetiva diferente daquela até então utilizada, ou seja, montar o perfil histórico dos clientes e buscar utilizações fora do padrão ao invés de focar em comportamento suspeito baseado apenas em perfil cadastral. Independente da tecnologia utilizada, esta proposta fortalece o conceito da política *Know Your Customer* (KYC) e torna-se ainda mais útil se aplicada de forma incremental e constante.

---

[1] http://ota.fas.org/ - escritório do Congresso Americano que funcionou de 1972 a 1995.

Em Tang & Yin (2005) os autores propõem outra extensão da SVM para analisar as transações dos clientes e detectar comportamento fora do padrão. É apresentada uma combinação de um kernel com melhoramentos do *Radial Basis Function* (RBF) Scholkopf *et al.* (2001) com definição de distâncias distintas e algoritmos SVM supervisionados e não-supervisionados. Uma vantagem desta abordagem é ela conseguir lidar com conjuntos de dados heterogêneos, porém, a avaliação de desempenho foi feita apenas com dados de simulação.

Combinar a técnica de *clustering* com *Multilayer Perceptron* (MLP) foi a proposta de Le-Khac *et al.* (2009). O algoritmo K-means, é utilizado para formação dos *clusters* e esta técnica baseia-se em duas características principais (fundo de investimento e investidor) que, em seguida, são usados como entrada do processo de formação de um MLP. Os resultados apresentados mostram que a sua abordagem é eficiente. No entanto, o número de características, o número de padrões de treinamento utilizados foi pequeno e isso pode afetar a precisão.

Outros métodos estatísticos foram propostos, como em Liu & Zhang (2010) que utiliza *scan statistics*, onde as transações realizadas num período de tempo são selecionadas aleatoriamente e agrupamentos incomuns são buscados. Adequado para trabalhos de auditoria em função da aleatoriedade adotada, no entanto, para um processo cotidiano de ABC nenhuma transação pode ser desprezada.

Em Le-Khac & Kechadi (2010) os autores apresentam um estudo de caso em que aplicam uma solução para geração de uma base de conhecimento, combinando técnicas de mineração de dados, *clustering* utilizando K-means, redes neurais e algoritmos genéticos para detectar padrões de BC. Analisando somente por este documento, a solução pode apresentar um baixo custo benefício, considerando a alta complexidade da implementação proposta.

Larik & Haider (2011) focam seu trabalho nas informações de débito e crédito realizadas pelos clientes de numa instituição financeira, visando identificar transações suspeitas. Propõem um novo algoritmo e um índice para avaliar e classificar as transações. Uma boa ideia, porém, limitada pela utilização somente das informações de débito e crédito.

### 2.2. Arquiteturas de Sistemas

Em Gao *et al.* (2006), uma arquitetura de sistema é definida utilizando um conjunto de agentes especializados, tais como: agentes de coleta de dados (sistemas internos e informações externas); agentes de monitoramento para acompanhamento do perfil cadastral do cliente e das transações realizadas; e um agente que emite relatórios e alertas sobre possíveis operações de BC. Apesar da boa proposta de arquitetura, o problema crucial do volume de análises submetidas ao analista humano não é enfrentando, aliás, pode até ser agravado com a automatização da fase de sinalização de transações suspeitas.

Outra abordagem baseada em agentes foi apresentada em Xuan & Pengzhu (2007). A arquitetura proposta é semelhante à descrita em Gao *et al.* (2006), porém, com evolução dos agentes incluindo as características de negociação, diagnóstico e autoaprendizagem. As características dos agentes é o ponto forte desta proposta, contudo, também encerra sua contribuição nas sinalizações para o analista humano.

Em Xu & Gao (2010) os autores propõem uma alteração na arquitetura apresentada anteriormente, incorporando as fases e técnicas de um modelo de tomada de decisão, aplicado em tempo real. A arquitetura do sistema proposto anteriormente foi melhorada, mas em termos de processo não houve avanços, persistindo a falta de apoio aos analistas humanos.

Rajput *et al.* (2014) utilizam ontologias e regras na criação de um sistema especialista para detecção de transações suspeitas de BC. Uma boa abordagem para a criação das regras a serem aplicadas, testada com um volume significativo de dados, contudo, a quantidade de 2% de transações suspeitas detectadas numa base real torna-se inviável num processo rotineiro que necessita da validação do especialista humano.

## 3. Estratégia de Combate a Fraudes

A política *Know Your Customer* (KYC) definida pelo Comitê de Basiléia[2] é um guia de melhores práticas no combate a fraudes, pois detalha os procedimentos a serem seguidos. Baseado nestes documentos e no *benchmark* realizado pela Hewlett-Packard (Laxman, 2014) foi criado um fluxo genérico de combate a fraudes ou burla no setor financeiro (Figura 1), quando realizadas em processos total ou parcialmente automatizados.

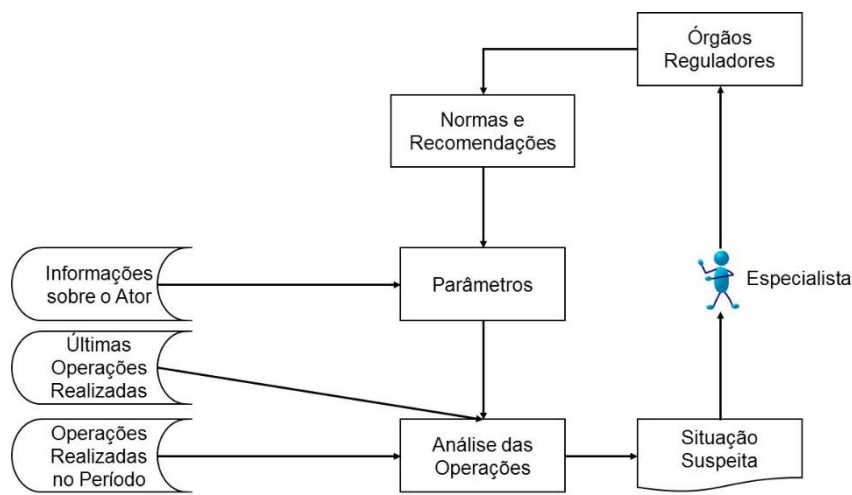

Figura 1 – Fluxo Genérico de Combate a Fraudes

O branqueamento de capitais (BC) é considerado um crime adjacente. Em Canas (2004) é explicado que "para haver Branqueamento teria de haver um crime anterior que proporcionasse ilicitamente ao seu autor proventos que posteriormente ele, ou outrem, pretendessem camuflar", na sua essência é possível afirmar que ele é uma instância do fluxo genérico de fraude. Dessa forma, a atividade de anti-branqueamento de capitais (ABC) pode ser vista como sendo o fluxo genérico mostrado na Figura 1, aplicado ao setor financeiro.

---

2 http://www.bis.org/publ/bcbs85.pdf e http://www.bis.org/publ/bcbs04a.htm

Um típico fluxo de ABC praticado pelas instituições financeiras foi mostrado e detalhado em Alexandre & Balsa (2015). Porém, mesmo sem a utilização de técnicas como mineração de processo (Norambuena & Zepeda, 2017), é fácil observar que o fluxo de ABC sendo tratado como uma instância do fluxo genérico de combate a fraudes é ineficiente, pois ele falha na identificação e sinalização das situações suspeitas. Casos não identificados impedem análise detalhada e a consequente geração de novas normas e recomendações, surgindo, dessa forma, o risco de uma falha sistêmica no processo geral de ABC.

A Figura 2 mostra uma nova versão do fluxo geral de combate a fraudes, cujo objetivo é mitigar o risco identificado. A criação de perfis dos atores participantes da atividade, baseados no histórico completo das operações realizadas; a substituição dos atuais parâmetros fixos por regras de produção, embasadas nestes perfis e nas normas e recomendações existentes; a eliminação da utilização de somente operações mais recentes; e o uso de inteligência em alguns pontos do processo eleva qualitativamente o nível na captura de operações suspeitas e, principalmente, da tomada de decisão pelo especialista (Pinto *et al.*, 2014).

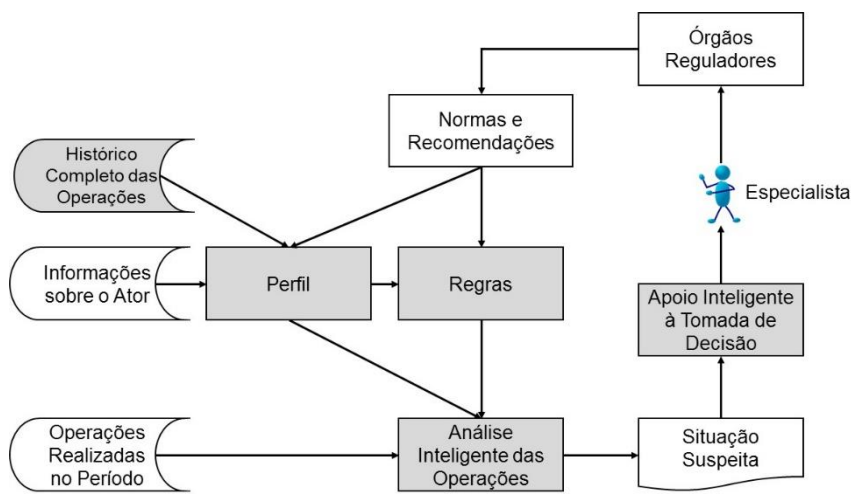

Figura 2 – Novo Fluxo Genérico de Combate a Fraudes

### 3.1. Abordagem baseada no comportamento transacional

Uma base com dados reais refletindo o comportamento transacional de clientes, possui uma significativa quantidade de atributos necessários para o controle e gestão do negócio envolvido, sendo que nem todos são relevantes para uma busca de transações suspeitas. A seleção dos atributos relevantes e a possível geração de novos atributos, reflete a importância da fase de pré-processamento, que será descrita na seção seguinte.

Os atributos selecionados ou gerados, precisam refletir as principais características do universo de dados onde a anomalia é buscada, no caso, devem refletir a base de transações e permitir identificar comportamentos suspeitos (Paula *et al.*, 2016).

Neste trabalho é proposta a criação de vários atributos, que agregam quantidades e segmentam características, com o objetivo de estabelecer um perfil do comportamento transacional, dentro de um ciclo temporal, para cada ator do processo. O ciclo temporal tem relação direta com a natureza do negócio envolvido, apresentando a duração máxima possível (trimestral, semestral, anual, etc.).

**3.2. Pré-processamento dos dados e formação dos perfis**

Neste trabalho são utilizados dados reais de um banco brasileiro, referente a dois anos de transações do produto contas correntes. As contas dos 5,2 milhões de correntistas deste banco recebem, em média, 85 milhões de transações anuais.

A análise da base de dados mostrou que menos de 10% dos clientes são constituídos por pessoas jurídicas (empresas comerciais, industrias, governos, etc.), no entanto, são responsáveis por mais de 90% dos valores totais envolvidos. Desta forma, visando uma melhor caracterização dos clientes, foi realizada a divisão da base de dados: uma somente com clientes singulares e outra com clientes tipo pessoa jurídica. Todo o procedimento descrito foi executado separadamente para cada base e está representado na Figura 3.

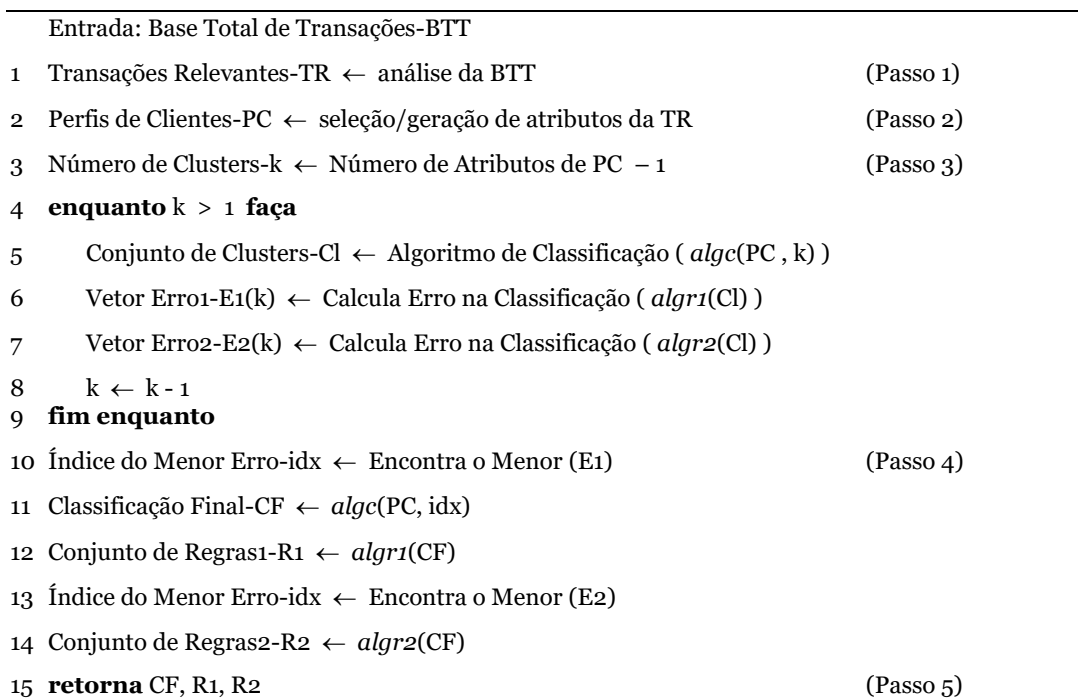

| | Entrada: Base Total de Transações-BTT | |
|---|---|---|
| 1 | Transações Relevantes-TR ← análise da BTT | (Passo 1) |
| 2 | Perfis de Clientes-PC ← seleção/geração de atributos da TR | (Passo 2) |
| 3 | Número de Clusters-k ← Número de Atributos de PC − 1 | (Passo 3) |
| 4 | **enquanto** k > 1 **faça** | |
| 5 |    Conjunto de Clusters-Cl ← Algoritmo de Classificação ( *algc*(PC , k) ) | |
| 6 |    Vetor Erro1-E1(k) ← Calcula Erro na Classificação ( *algr1*(Cl) ) | |
| 7 |    Vetor Erro2-E2(k) ← Calcula Erro na Classificação ( *algr2*(Cl) ) | |
| 8 |    k ← k - 1 | |
| 9 | **fim enquanto** | |
| 10 | Índice do Menor Erro-idx ← Encontra o Menor (E1) | (Passo 4) |
| 11 | Classificação Final-CF ← *algc*(PC, idx) | |
| 12 | Conjunto de Regras1-R1 ← *algr1*(CF) | |
| 13 | Índice do Menor Erro-idx ← Encontra o Menor (E2) | |
| 14 | Conjunto de Regras2-R2 ← *algr2*(CF) | |
| 15 | **retorna** CF, R1, R2 | (Passo 5) |

Figura 3 – Algoritmo do Processo de Aprendizagem

De cada base foram excluídas as transações cujas características impedem a prática de branqueamento de capitais (tarifas, comissões, juros, impostos, etc.) (Figura 3 - Passo 1). A soma final das bases resultou em 35 milhões de transações relevantes.

Um ano de transações foi utilizado para a geração do perfil de comportamento transacional, ou seja, para cada cliente presente na base de transações foram agrupadas informações tais como: idade da conta; quantidade de movimentos gerados e de serviços utilizados; percentual de transferência de recursos para outros bancos e para contas do próprio banco; além da quantidade de movimentos agregados em 6 faixas de valores. O 12º atributo, denominado percentual de débito, representa, de forma ponderada no período analisado, o tempo que o recurso permaneceu na conta do cliente (Alexandre & Balsa, 2016) (Figura 3 - Passo 2).

### 3.3. Aprendizagem e geração de regras

A tabela de perfis dos clientes ativos no ano analisado ficou com 2,4 milhões de linhas, cada linha representando univocamente a tripla cliente, agência e conta. Esta tabela foi utilizada num processo de aprendizado indutivo não-supervisionado com *clustering*, para formação de grupos de clientes com características semelhantes e mutuamente exclusivos. O algoritmo K-means (*algc*) foi utilizado para classificação e os algoritmos PART e J48 (*algr1* e *algr2*) para geração das regras de produção, executados 11 vezes (número de atributos menos 1) (Figura 3 - Passo 3).

O modelo de *cluster* utilizado mostrou-se adequado ao permitir a interpretação dos agrupamentos sem a utilização de sofisticados esquemas de visualização (Castillo-Rojas *et al.*, 2017). Dois conjuntos de regras, um de cada algoritmo utilizado, com a menor quantidade de instâncias incorretamente classificadas foram selecionados (Figura 3 - Passo 4) e conjuntamente com os clusters que originaram essas regras representam o resultado final do processo (Figura 3 - Passo 5).

## 4. Abordagem Conservadora com Relação a Risco

A análise dos clusters gerados, para os dois segmentos de clientes, permitiu a identificação de características tais como: grande movimentação de valores elevados, com transferência integral para outras instituições financeiras (risco 3); ou movimentação de valores próximos do limite de comunicação aos órgãos reguladores (risco 2). Desta análise resultou a classificação mostrada na Tabela 1.

Com esta classificação é possível definir uma melhor estratégia, oferecendo tratamento diferenciado aos grupos de clientes, conforme seu nível de risco. Apesar do ótimo nível de acerto obtido na avaliação das regras geradas, em torno de 99% para ambos os segmentos de clientes, um por cento de erro representa mais de 26 mil transações e não podem ser desprezadas.

A matriz de confusão gerada pelos algoritmos identificou as regras que por serem aplicáveis a dois ou mais grupos de clientes compõem o 1% de erro mencionado. Ou seja, regras classificam clientes como pertencentes a mais de um perfil. A decisão foi então reclassificar os perfis que não representam risco ou têm baixo risco (perfis 1, 2 e 3), conforme mostra a

Tabela 1. Dessa forma, a transação que pertencer a um desses 3 grupos será reclassificada, somente para efeito de análise, para uma regra nos perfis de risco mais elevado que satisfaça a condição.

Por exemplo, na base de dados utilizada para *data mining*, 33 regras classificam os clientes pessoas singulares como pertencentes aos perfis de risco 2 e 3, correspondendo a 1,85% do total. Contudo, estas regras também classificam 0,06% de clientes originalmente pertencentes ao perfil padrão. A reclassificação consiste em, durante o processo de busca por transação suspeita, considerar esses clientes padrão como pertencentes aos grupos de risco, sem modificar a classificação original.

### 4.1. Criação de regras normativas e baseadas no comportamento

A classificação dos perfis também permite a criação de regras especificas, quer sejam baseadas nos normativos vigentes, quer sejam inspiradas no comportamento transacional. Conforme já mencionado, este trabalho utilizou um ano de informações para geração dos perfis, obtendo totais mensais e permitindo selecionar valores máximos dentro do ano para cada atributo relevante.

Foi estabelecido que a busca por transações suspeitas retroagirá sempre um mês a partir da data solicitada para análise, desta forma será sempre utilizado o comportamento de 1 mês de transações para efeito de comparação com os perfis.

Para os perfis de risco, as regras utilizarão sempre cálculos em torno do maior valor mensal encontrado para cada atributo e, para os demais perfis, será utilizado como limite o valor total anual, conforme consta da Tabela 1.

Tabela 1 – Classificação dos perfis gerados

| Perfil | Pessoa Singular | Outro Tipo de Cliente | Reclassifica Perfil | Valor Limite |
|---|---|---|---|---|
| 1. *Baixa Utilização* | Cluster4 | Cluster4 | Sim | Total anual |
| 2. *Cliente Padrão* | Cluster2 | Cluster1 | Sim | Total anual |
| 3. *Risco 1 (baixo)* | Cluster1 | - | Sim | Máximo mensal |
| 4. *Risco 2 (médio)* | Cluster5 | Cluster3 | Não | Máximo mensal |
| 5. *Risco 3 (alto)* | Cluster3 | Cluster2 | Não | Máximo mensal |

O banco de regras criado deve ser o mais estável possível, posto que se transformará em crenças a serem utilizadas pelos agentes, possibilitando tomada de decisão consistentes e coerentes. No entanto, é possível que a instituição financeira utilizadora do sistema deseje, numa determinada análise, ser mais rigorosa quanto aos valores limites, o que levaria, inevitavelmente, a modificação nas regras. Para contornar esta situação, foi parametrizado um percentual intitulado Margem Adicional de Risco (MAR), cuja atribuição ficará a critério da instituição utilizadora e que aplicado sobre os valores limites reduzem-no pelo percentual indicado. Nos testes realizados para este artigo, referido percentual foi mantido em zero.

A Figura 4 mostra que quando a MAR é maior que zero (o que está indicado em setembro, e no Total) os perfis de risco 2 e 3 têm seu limite reduzido pelo percentual indicado. O mesmo ocorre com os demais perfis, sendo que estes em relação ao valor total anual. O perfil de risco 1, por ser perfil de alerta, não é afetado pela MAR. Quando a MAR é igual a zero os perfis de risco utilizam o mesmo valor limite e os demais perfis utilizam o valor total anual, sem reduções.

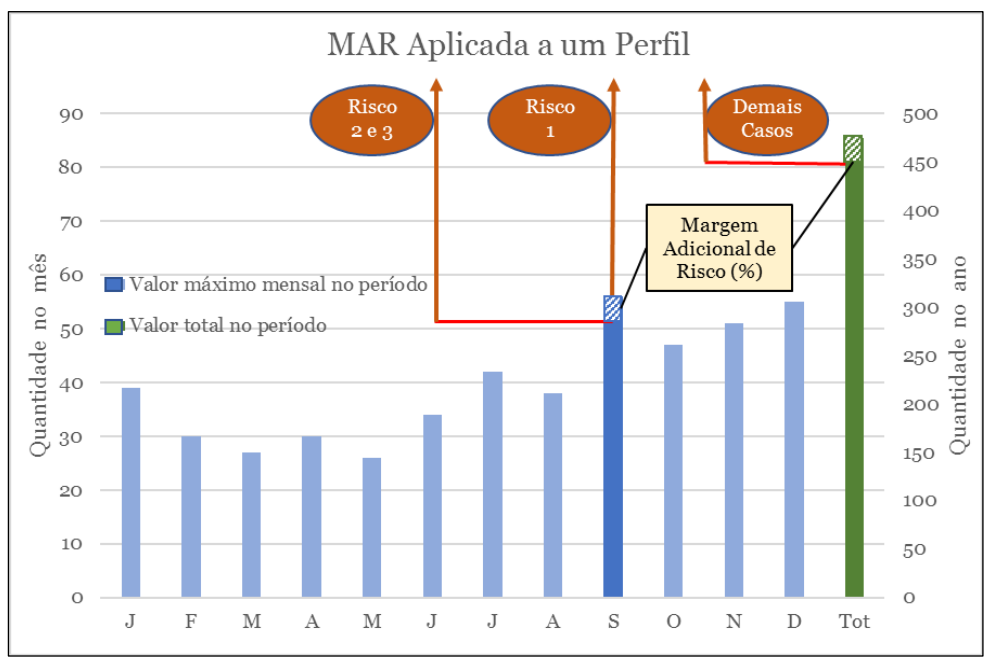

Figura 4 – Aplicação da Margem Adicional de Risco

## 5. Modelo dos Agentes

A maioria dos sistemas existentes atualmente para suportar o processo de anti-branqueamento de capitais (ABC), concentra-se na fase do processo referente a captura de transações suspeitas, transferindo para o Analista de ABC a tarefa de comprovação da suspeição. Além disso, boa parte dessas soluções são fortemente baseadas em parametrizações que, basicamente, aplicam os normativos vigentes.

O desempenho desses sistemas está refletido no resultado da pesquisa realizada em 2014 com 317 profissionais especialistas em ABC e técnicos de conformidade, de instituições financeiras de 48 países, mostrando que apenas 58% acreditam que os sistemas existentes em suas organizações são capazes de monitorar adequadamente as transações ocorridas nas diversas linhas de negócios[3].

Esta situação motiva este trabalho na proposição de uma solução que integra técnicas de *data mining* a um sistema multiagente com o propósito de capturar transações suspeitas de branqueamento de capitais e auxiliar o Analista de ABC na tomada de decisão. Na modelagem do sistema foi utilizada a metodologia *Prometheus* (Padgham & Winikoff, 2004). A implementação está sendo realizada no *framework* JaCaMo (Boissier *et al.*, 2013) que utiliza a linguagem Jason (Bordini, Hübner & Wooldridge, 2007).

A Figura 5 mostra, de forma simplificada, um dos diagramas da *Prometheus Design Tool*[4](PDT), com a arquitetura final do sistema. O diagrama ressalta os agentes, as bases de dados externas (DB), as bases de conhecimento (KB) e a interação dos agentes com o ambiente e os Analistas de ABC. O grupo de agentes responsáveis pela captura de transação suspeita (CTS) que são especializados por produto da linha de negócios de uma instituição financeiro (contas correntes, câmbio, fundos de investimento, empréstimos, etc.). Esta especialização oferece duas grandes vantagens: a primeira é que cada agente pode ser aprimorado com as especificidades do produto e dos perfis dos clientes, que podem mudar em cada produto; a segunda diz respeito aos princípios de manutenibilidade e de escalabilidade, ou seja, um produto vigente pode ser descontinuado (morte de um CTS) e um novo produto pode ser criado (novo CTS) sem que isso interfira no funcionamento dos demais agentes.

Na análise das transações os CTS utilizam as regras de produção vigentes, geradas com base nos perfis dos clientes, nos normativos sobre ABC e nas normas internas da instituição financeira. Atuam em duas modalidades: busca por transação ou busca por cliente. Na busca por transação a base histórica de transações é analisada integralmente, dentro do período informado, enquanto na busca por cliente, somente as transações relacionadas ao cliente informado são analisadas.

Um agente é responsável pela gerencia da captura de transações (GCT), ele pode receber uma solicitação externa de análise e comandar para execução pelo CTS especialista ou comandá-la autonomamente. Ao receber a informação de um CTS de que uma transação suspeita foi identificada, comanda uma análise na modalidade busca por cliente para os demais CTSs responsáveis por produtos que constem do perfil do referido cliente. Somente o GCT tem conhecimento de quantos e quais são os CTSs existentes. Após todos os CTSs acionados terem enviado mensagens de resposta ao comando de análise, o GCT comunica ao agente auxilia processo decisório (APD) a existência de transações suspeitas.

O agente APD é responsável por aprofundar a análise e decidir sobre a suspeição da transação, confirmando-a ou não. A utilização de conhecimento específico sobre o produto e de uma matriz de decisão baseada no histórico de decisões tomadas, permitem ao agente decidir sobre a suspeição ou, quando não é possível chegar a uma decisão, sinalizar para o

---

[3] https://home.kpmg.com/xx/en/home/insights/2014/01/global-anti-money-laundering-survey.html

[4] https://sites.google.com/site/rmitagents/software/prometheusPDT

Analista de ABC sobre a existência de um caso complexo. As decisões tomadas pelo agente, bem como aquelas informadas pelo Analista, são guardadas no histórico de decisões e são utilizadas no aprendizado para evolução da matriz de decisão. Este agente também sugere alterações na matriz de decisão e pela atualização desta base de conhecimento, após as sugestões serem validadas.

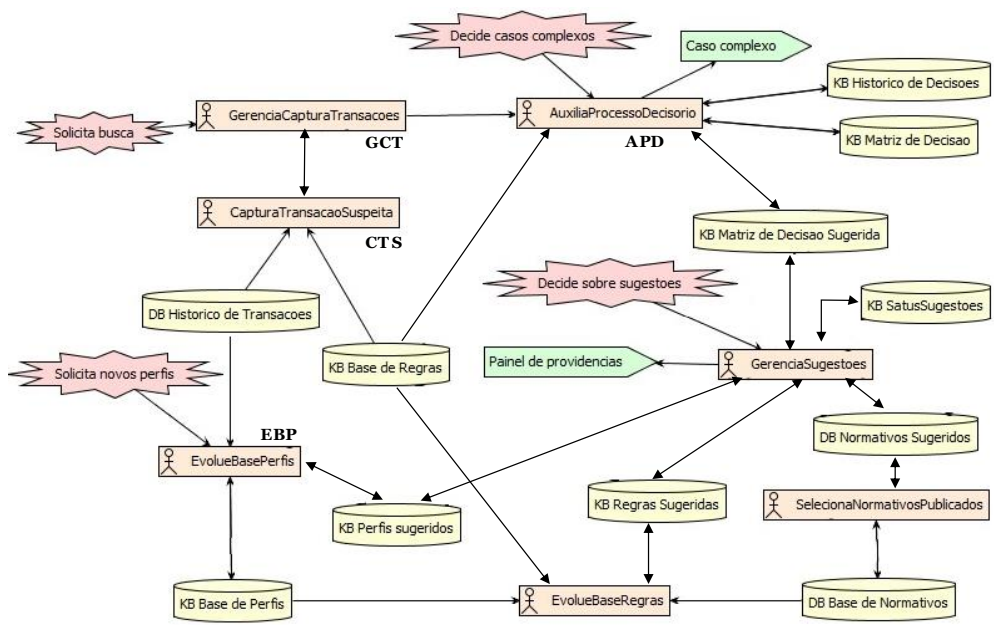

Figura 5 – Arquitetura do Sistema Proposto

O agente evolui base de perfis (EBP) atua na análise do histórico de transações para geração de perfis de clientes e posterior comparação com a base de perfis existente. Este processo pode ser acionado por uma solicitação do usuário ou de forma autônoma pelo agente. Os possíveis novos perfis surgidos são sugeridos para o Analista de ABC, para análise. O agente EBP atualiza a base de perfis com os perfis que forem validados.

As evoluções das bases de conhecimento e o aprendizado previstos no sistema visam, primordialmente, mitigar o risco da ocorrência de falso positivo e/ou falso negativo, existente em sistema baseados unicamente em um conjunto de regras e padrões de comportamento (Gao *et al.*, 2006) e (Le-Khac & Kechadi, 2010).

## 6. Análise dos Resultados

Os dados reais utilizados neste trabalho referem-se a 2 anos, com 30,5 milhões e 35,2 milhões de transações relevantes, respectivamente. Os perfis de comportamento transacionais dos clientes foram gerados a partir dos dados do primeiro ano, constituindo a base de referência. A busca por transações suspeitas foi realizada em um mês de transações relevantes do segundo ano, resultando em 2,6 milhões de transações. Sobre estas transações foram gerados 516.942 perfis de comportamento transacionais dos clientes no período.

O processo de busca por transações suspeitas, implementado até o momento, pode ser dividido em 3 fases: a reclassificação dos perfis ou ajuste da matriz de confusão; a captura das transações suspeitas; e a análise das transações capturadas. A Figura 6 mostra um resumo com informações do ambiente e a quantidade de transações suspeitas.

```
== Processo de captura iniciado ... 20:16:30

444.819 perfis selecionados      Margem Adicional de Risco: 0%

Fase 1 - Reclassificação dos Perfis (ajuste matriz de confusão)
 Aplica regras geradas no processo de aprendizagem
 (1) Pessoas Singulares/Físicas   49 regras    Versão: 30112016.01
 (2) Outros Tipos de Pessoa       55 regras    Versão: 30112016.01
 ------------------------ (1) ------------ (2) ----
 Perfis------------Original Ajuste---Original Ajuste
 Baixa utilização   60.241      0      4.552     0
 Padrão            295.424   -413      6.853    -5
 Alerta              4.768      0          0     0
 Risco               4.886    +39      8.472     0
 Alto Risco         21.305   +374     38.318    +5
 Erro                    0      0          0     0
 Total de Perfis   386.624            58.195

Fase 2 - Captura transações suspeitas (aplica regras)
  5 regras baseadas em normativos         versão: 02032017.01
 20 regras baseadas nos perfis            versão: 02032017.01

 182 perfis suspeitos
 0.0409% dos perfis analisados

== Processo de captura concluido ... 22:23:37
```

Figura 6 – Resultado Fases 1 e 2

O sistema utiliza como crenças um conjunto de 129 regras, bem como a classificação dos clientes nos perfis indicados. Essas crenças são reavaliadas uma única vez, ocasião em que 418 perfis foram reclassificados, saindo do perfil Padrão para perfis de risco, somente nesta análise. No final 182 perfis são indicados como suspeitos e suas transações precisam ser analisadas.

O resultado da fase 3 do processo é mostrado na Figura 7. A distribuição dos suspeitos por perfil, com seu respectivo percentual, alerta para a concentração de 52% dos suspeitos no perfil de risco 3 (Alto Risco). Com o objetivo de auxiliar o especialista na análise das transações, os suspeitos são agrupados pela regra acionada na sua captura (mais de uma regra pode ser acionada para o mesmo perfil).

```
== Processo de análise iniciado ... 23:14:32

Fase 3 - Análise das transações sinalizadas
  Captura realizada em: 23/3/2017  Analisados: 444.819 perfis   Margem Adicional de Risco(MAR): 0%

Distribuição dos suspeitos por perfil:
   Baixa utilizacao -    8      4%
   Padrao           -   32     18%
   Alerta           -    1      1%
   Risco            -   47     26%
   Alto Risco       -   94     52%
   Total Suspeitos  -  182

Foram ativadas 15 regras, que ocorreram 184 vezes:
   BCXX2016002 -   9       PCXX2016003 -  13       PCXX2016014 -   1
   BCXX2016003 -  20       PCXX2016004 -   1       PCXX2016015 -  48
   BCXX2016004 -   4       PCXX2016005 -   4       PCXX2016017 -   4
   PCXX2016001 -   1       PCXX2016006 -  18       PCXX2016018 -  15
   PCXX2016002 -  27       PCXX2016012 -   6       PCXX2016020 -  16

Analise de 182 suspeitos, iniciando no número 1.  Regra específica(["Todas"])
Suspeito: 1 / 25     Cliente-█████   Agencia-██  Conta-█████  (21 anos)  Status-risco3 (K/ot)
(Atributo - total anual / valor máximo no mês / valor no mês)
 Serv    -    46 /    9 /    4      Movi   - 1776 /  830 /    8
 FxVlr1  -    19 /   11 /    0      FxVlr4 -  222 /  117 /    0
 FxVlr2  -   101 /   74 /    0      FxVlr5 -  109 /   60 /    4
 FxVlr3  -  1312 /  709 /    0      FxVlr6 -   13 /    6 /    4
 PctDEB  - 111.11/ 111.11/ 92.31     PctTED -    0/    0/    0
 PctDOC  -     0/      0/    0
 BCXX2016003-Queda acentuada na movimentação em perfil de alta movimentação (CC.3542.1.IV.f)
-----
Suspeito: 74 / 137149  Cliente-█████   Agencia-██  Conta-█████  (11 anos)  Status-risco3 (J/ot)
(Atributo - total anual / valor máximo no mês / valor no mês)
 Serv    -    31 /    4 /    4      Movi   -   92 /   15 /   12
 FxVlr1  -     0 /    0 /    0      FxVlr4 -    0 /    0 /    0
 FxVlr2  -     0 /    0 /    0      FxVlr5 -   39 /    4 /    2
 FxVlr3  -     0 /    0 /    0      FxVlr6 -   53 /   12 /   10
 PctDEB  - 92.47/ 98.64/ 96.03       PctTED - 19.79/ 78.9/ 83.5
 PctDOC  -     0/     0/    0
 PCXX2016006-PctTED > (MAR x máximo mensal) e FxVlrs4-6 >= 90% do movimento mensal(Perfis de risco)
```

Figura 7 – Resultado Fase 3

Em função do sigilo legal envolvido nas informações, a Figura 7 mostra apenas dois perfis com parte dos detalhes envolvidos, devidamente descaracterizados. No entanto, é possível observar o texto de algumas regras que ressaltam a relevância das suspeitas, como por exemplo a regra legal BCXX2016003, originada no normativo do Banco Central Brasileiro que sinaliza uma queda acentuada na movimentação de um perfil que, normalmente, tem alta movimentação. Os normativos dos bancos centrais dos países, geralmente, recomendam investigar esta mudança de comportamento pois pode representar a parada de um período no qual ocorreu o branqueamento de capital. O percentual de 0,0409% de suspeitos, para um mês de transações, mostra-se exequível para a análise humana.

O sistema atualmente em utilização na instituição financeira fornecedora dos dados é fortemente baseado em valores limites pré-fixados e informações cadastrais, tipo renda/faturamento e atividade econômica. Assim sendo, as transações suspeitas agora indicadas incluem apenas algumas das suspeitas efetivamente identificadas e comunicadas aos órgãos de controle. As suspeitas baseadas no comportamento transacional dos clientes estão em análise visando atestar seu grau de precisão.

## 7. Conclusões

É crescente a preocupação da indústria financeira e dos governos com o crime de branqueamento de capitais, quer seja pelos recursos perdidos, quer seja pelas consequências deste crime que, normalmente, financiam outros crimes.

Os sistemas atualmente em utilização pelas instituições financeiras são fortemente baseados em informações cadastrais, a proposta aqui apresentada analisa o comportamento transacional dos clientes, estabelece um perfil e utiliza-o como balizador para comportamento futuro. Os resultados obtidos mostram a viabilidade de utilização sistemática e estabelece nova frente de combate a este crime.

A verificação real das suspeitas sinalizadas estão em andamento e o próximo passo é fazer com que o agente de auxílio ao processo decisório aprenda com as decisões tomadas pela Analista de Anti-Branqueamento de Capitais e passe a sinalizar somente os casos inéditos.

## Referências